\documentclass[aps, amsfonts, nofootinbib,notitlepage, preprintnumbers]{revtex4-1}
\usepackage{hyperref}
\hypersetup{
colorlinks=true,
linkcolor=red,
citecolor=blue,
urlcolor=blue}
\usepackage{epsf,epsfig}
\usepackage{amscd}
\usepackage{amsmath}
\usepackage{subfig}
\usepackage{graphicx}
\usepackage[countmax]{subfloat}

\input xy
\xyoption{all} 

\newcommand{\figref}[1]{Fig.~\ref{#1}}

\newcommand{\secref}[1]{Section~\ref{#1}}
\newcommand{\reference}[1]{Ref.~\cite{#1}}

\newcommand{\oneover}[1]{\ensuremath{\frac{1}{#1}}}
\newcommand{\half}{\oneover{2}}

\newcommand{\PiT}{\ensuremath{\Pi^\perp}}
\newcommand{\Tcoulomb}{\ensuremath{T_\text{Coulomb}}}
\newcommand{\Tbec}{\ensuremath{T_\text{BEC}}}

\newcommand{\nn}{\nonumber}
\newcommand{\Lag}{{\mathcal L}}
\newcommand{\Dv}{{\mathbf{D}}}
\newcommand{\Ab}{{\mathbf{A}}}

\begin{document}

\title{Thermodynamics of nuclear condensates and phase transitions in white dwarfs} 
\author{Paulo F.~Bedaque\footnote{{\tt bedaque@umd.edu}}} 
\author{Evan Berkowitz\footnote{{\tt evanb@umd.edu}}}
\author{Srimoyee Sen\footnote{{\tt srimoyee@umd.edu}}}
\affiliation{Maryland Center for Fundamental Physics,\\ 
Department of Physics,\\ 
University of Maryland, College Park, MD USA}

\preprint{UM-DOE/ER/40762-522}

\begin{abstract}
	
	We study the thermodynamics of  helium at densities relevant for white dwarf physics. We find evidence that, as the temperature is increased, there is first a first order transition between two superconducting phases followed by a second order transition to the normal state. These transitions occur, for realistic densities, at temperatures below the crystallization temperature and the crystalline state is likely to remain as the true ground state of the system. The calculations are performed with a screening but non-dynamical electron background and we comment on the impact of this and other approximations to our result.

\end{abstract}
\maketitle

\section{Introduction}

Helium white dwarfs (He WDs) are astrophysical objects which are composed predominantly of helium nuclei and degenerate electrons.  At typical WD densities, the nuclei are much closer together than typical atomic sizes but are still widely separated compared to typical nuclear sizes.  It has long been known that as WDs cool the nuclei crystallize, locked into position by their mutual Coulomb interactions\cite{abr60,kir60,sal60}.  Recently, it was pointed out that in He WDs the temperature at which the helium nuclei form a Bose-Einstein condensate (BEC) might be higher than the crystallization temperature and an intermediate superconducting phase may exist between the plasma and the crystal phases \cite{Gabadadze:2008mx,Gabadadze:2007si,Gabadadze:2009jb,Gabadadze:2009dz,PhysRevLett.66.2915,Ashcroft:kx}. 
In this phase, it is the ions that are superconducting; the electrons form an ordinary Fermi liquid. The low temperature properties of this phase are dominated by the physics of an unusual ``phonon" excitation \cite{Bedaque:2011hs} and leads to a very small specific heat and enhanced neutrino emission \cite{Bedaque:2012mr}, with possible consequences for the cooling of He WDs \cite{Benvenuto:2011fj}.  A similar phase could also exist in a deuterium layer in brown dwarfs \cite{Berezhiani:2010db} and be relevant for inertial confinement  \cite{PhysRevLett.78.483,Silva:1997fk,Jeanloz29052007} as well as other kinds of experiments \cite{Badiei200970,Andersson20093067} where high densities are also achieved .

It is guaranteed that at large enough densities there will be a range of temperatures where the BEC can exist while the Coulomb crystal cannot.  This can be understood by simple scaling arguments:  a BEC should form when the thermal de Broglie wavelength $\sqrt{2\pi/MT}$ (here $M$ is the ion mass and $T$ is the temperature) becomes comparable to the interparticle spacing $l$, so that the condensation temperature should scale as $\Tbec\sim 1/Ml^{2}$.  A Coulomb crystal should melt when the thermal energy is comparable to the nearest-neighbor interaction, so that $\Tcoulomb\sim Z^{2}\alpha/l$, where $Z$ is the atomic number of the crystallized nuclei and $\alpha=e^{2}/4\pi\approx 1/137$, where $e$ is the size of the electron charge.  Since the number density $n\sim l^{-3}$, $\Tbec\sim n^{2/3}/M$ while $\Tcoulomb\sim n^{1/3}$.  Thus, at very high density, the crystallization temperature is markedly lower than the condensation temperature, and for intermediate temperatures, the system should be a BEC.

The natural question is: are astrophysical densities in this interesting regime?  To answer this quantitative question, one needs to know the numerical coefficients that specify these critical temperatures.  Because the condensation temperature scales inversely with the ion mass, the density at which $\Tbec=\Tcoulomb$ and beyond which \mbox{$\Tbec>\Tcoulomb$} is smaller for lighter nuclei.  Thus, if a nuclear condensate forms in WDs, it should be most easily established in He WDs and not carbon-oxygen WDs.

Detailed studies have determined the crystallization temperature to be $\Tcoulomb \sim (Ze)^{2}/180 l$ \cite{1975ApJ...200..306L,PhysRevA.21.2087,1993ApJ...414..695C}, meaning $\Tcoulomb\sim (a_{0}/l) 7000$K, where $a_{0}$ is the Bohr radius.  There are various suggestions for the proportionality constant in \Tbec.  Simply equating the de Broglie wavelength to the interparticle spacing suggests $\Tbec = 2\pi/Ml^{2}\approx 6.2/Ml^{2}$.  A free Bose gas has $\Tbec=T_{c}^{(0)}\equiv 2\pi (4\pi\zeta(3/2)/3)^{-2/3} / Ml^{2}\approx 1.27/Ml^{2}$, where $\zeta$ is the Riemman zeta function.  The temperature \Tbec\ is expected to go up when one considers repulsive interactions \cite{Huang:1999zz}.  A slightly more detailed estimate (see \reference{Gabadadze:2009jb}) suggests $\Tbec=4\pi^{2}/3Ml^{2}\approx 13.2/Ml^{2}$, which is qualitatively supported by the numerical calculations in \reference{Rosen:2010es}.  It is the object of this paper  to make a reliable estimate of \Tbec. 

A BEC composed of nuclei (and not whole atoms) with a background of degenerate electrons is a novel system with rich phenomenology.  Because the condensed nuclei are charged, the substance is electrically superconducting.  The electrons provide a neutralizing electric charge, and additionally the dynamical response of these electrons implies an unusual gapless quasiparticle\cite{Bedaque:2011hs}.  These quasiparticles imbue the substance with a very small specific heat \cite{Bedaque:2011hs}.  Moreover, these quasiparticles can annihilate into neutrinos, and the power emitted per unit volume scales like $T^{11}$, so that the phenomenological relevance of this annihilation for He WD cooling depends strongly on the critical temperature of the nuclear condensate, with higher temperatures corresponding to more relevant neutrino emission\cite{Bedaque:2012mr}.  These considerations motivate the detailed study of the thermodynamics of such a nuclear condensate.

 For calculational simplicity, we will work in the regime of stiff electrons, so that they simply screen the Coulomb interaction with a screening mass.  In this light, our investigation may be seen as an investigation of nonrelativistic charged spin-0 bosons interacting via a screened Coulomb (that is, Yukawa) interaction.  Surprisingly, we find that the system is significantly more complex than expected and that we can merely set an upper bound for the first-order transition temperature: $\Tbec<T_{c}^{(0)}$.  We conjecture that this low first-order transition temperature is accompanied by an unforeseen second-order transition at $T_{c}^{(0)}$.

This paper is organized as follows:  in \secref{sec:model} we discuss this model in detail and calculate its one-loop effective potential.  In \secref{sec:phase-diagram} we establish the phase diagram described by this model and investigate its properties analytically and numerically.  In \secref{sec:global} we demonstrate that the condensed phase is globally disfavored anywhere the usual uncondensed phase exists, and to resolve this puzzle conjecture that the phase transitions that this system undergoes are more complicated than previously appreciated.  Finally, in \secref{sec:conclusion} we make some remarks about the phenomenological relevance of nuclear condensates and discuss priorities for more deeply understanding this model.

\section{Model, quasiparticles and effective potential}\label{sec:model}

At the densities we are considering our system can be described by the (Euclidean space) Lagrangian

\begin{equation}\label{eq:L_initial}
\Lag = \psi^\dagger \left( D_0- \mu - \frac{\Dv^2}{2M}\right)\psi + \frac{1}{4}F_{\mu\nu}F_{\mu\nu}+\mathcal{L}_{gauge} +\bar\eta (D_\mu\gamma_\mu+m+\mu_e \gamma_0)\eta,
\end{equation} 
where $\psi$ is the spin-0 helium nucleus field with charge $Ze=2e$ so that $D_\mu\psi = \partial_\mu\psi - i Ze A_\mu\psi$, $\eta$ represents the usual electron, with $D_{\mu}\eta=\partial_{\mu}\eta + i e A_{\mu}\eta$, $A_\mu$ is the photon field, $M$ ($m$) is the helium nucleus (electron) mass and $\mu$ ($\mu_e$) is the chemical potential for the nuclei (electrons). $\mathcal{L}_{gauge}$ is the gauge-fixing action required for perturbative calculations.

The Lagrangian in \eqref{eq:L_initial} does not contain the nuclear force between ions.  We omit this force because in the regime we are considering the Coulomb repulsion prevents two nuclei from approaching one another to distances comparable to the nuclear size and, consequently, the nuclear force between them is inoperative: the nuclei interact with each other (and with electrons) only through the electromagnetic force.  At the densities we consider ($n < 10^{8} g/cm^3 $) the nuclei are non-relativistic while the electrons may or may not be relativistic. A chemical potential for the electron is included and chosen so that the charge density of electrons equals that of the nuclei, ensuring charge neutrality.

Despite its apparent simplicity the action above describes a tremendous array of phenomena; this is not surprising giving the number of parameters present. We will concentrate on the regime described in the introduction where three different small parameters can be identified, namely i) $\alpha m l = l/a_0$, the ratio between the particle distances and the Bohr radius, ii) the fine structure constant $\alpha$ and iii) the mass ratio $m/M$. In order to make progress we will attempt a calculation that captures the leading order effect on these three parameters. Sometimes, however, certain effects will be proportional to ratios of these parameters. In these circumstances we will consider their numerical value to decide which terms to neglect. For instance, we count $m/(\alpha M)\approx 10^{-2}$ as a small parameter.

The first step is to integrate out the electrons. The result is, in general, a complicated  non-local action for nuclei and photons only. Later, we will use only the part of the action quadratic in the fields. So, at leading order in $\alpha$ we have
\begin{equation}
\mathcal{L}_{\psi A} = \psi^\dagger \left( D_0- \mu - \frac{\Dv^2}{2M}\right)\psi + \frac{1}{4}F_{\mu\nu}F_{\mu\nu} + \Lag_{gauge} +\frac{1}{2}A_\mu   \Pi_{\mu\nu}A_\nu -  (eA_0+\mu_{e}) n.
\end{equation} The quadratic term in $A_\mu$ is the one-loop photon polarization tensor due to the electrons. Higher loop corrections are suppressed by powers of $\alpha ml = l/a_0$ and are small for  the dense electron plasmas considered here \cite{Fetter:1971fk}.

In the density and temperature regime considered here, the electrons are degenerate and we can use the $T=0$ form of the polarization tensor 
\begin{equation}
\Pi_{\mu\nu} =
\begin{pmatrix}
\Pi  &   -\frac{p_ip_0}{\mathbf{p}^2}\Pi \\
  -\frac{p_j p_0}{\mathbf{p}^2}\Pi  &      \frac{p_i p_j p_0^2}{\mathbf{p}^4}\Pi +
  (p_i p_j-\delta_{ij}\mathbf{p}^2)\PiT
\end{pmatrix}
\end{equation} 
where $\Pi$ and $\PiT$ are functions of $p_0, \mathbf{p}$ . The form of the polarization tensor and the fact that it is determined by two functions follow from the Ward identity
$p_\mu \Pi_{\mu\nu}=0$. $\PiT$ will play no role in what follows but $\Pi$ will be essential for our discussion. It can be written as $\Pi(p_0, \mathbf{p}) = m_s^2 f(p_0 m/k_F^2, \mathbf{p}/k_F)$ where $f\rightarrow 1$ at small $p_0 m/k_F^2, \mathbf{p}/k_F$  \cite{Fetter:1971fk}. The zero momentum value of $\Pi$ describes the static screening of Coulomb forces by the cold electron gas. For certain values of the momentum $\Pi$ also has a small imaginary part. In this paper we will neglect both the imaginary part and the momentum dependence of $\Pi$, effectively working with a model where the electrons provide a negative charge background canceling the ion charge and  screening for the Coulomb force. In reality the electron gas leads to the ``Friedel oscillations" in the screened Coulomb force that will be considered in a further publication. Our calculations are thus applicable to a model of spinless bosons interacting through a screened Coulomb (Yukawa) potential. The effect of the full momentum dependence of $\Pi$ and the Friedel oscillations on the thermodynamics will be left for a later publication.

We are interested in the possibility that nuclei condense, that is, that the condensate $\langle \psi \rangle=v$ be non-zero, breaking the electromagnetic $U(1)$ symmetry spontaneously. Whether this happens or not can be decided by minimizing the effective potential $V(v)$\footnote{We are assuming that translation symmetry is not spontaneously broken and $v$ is position independent.}. For this purpose we compute now the one-loop effective potential by using standard methods \cite{Peskin:1995ev}. First, we split the nuclear field into a classical part $v$ and a fluctuating piece $\chi$ as $\psi=v+\chi_R + i \chi_I$. Then we  expand the action to quadratic order in the fields $A_\mu, \chi, \chi^\dagger$ and perform the gaussian path integral. The effective potential is given by

\begin{align}
e^{-\beta\int d^3r V(v)} 
&= \int D\chi D\chi^\dagger DA_\mu\ e^{-S_{quad}}\nn\\
&=\left({\rm det}\ S_{quad} \right)^{-1/2}\nn\\
&= e^{-\frac{1}{2}{\rm tr}\ln S_{quad}},
\end{align} where $\beta=1/T$ is the inverse temperature. 

The fields $\chi, \chi^\dagger$ and $A_\mu$ mix in the quadratic part of the action. The unitary gauge-fixing used in \reference{Bedaque:2011hs},
\begin{equation}
\mathcal{L}_{gauge} = -\frac{1}{2\xi}\left(
\nabla.\Ab - \frac{2M}{ev}\partial_0 \chi_R - \frac{\xi Zev^2}{M} \chi_I
\right)^2,
\end{equation} followed by the limit $\xi\rightarrow \infty$ allows us to decouple the fields at quadratic order and therefore simplify many calculations. However, the use of the unitary gauge (and $R_\xi$ gauges in general) is known to be problematic at finite temperature \cite{Dolan:1973qd}. The gauge fixing condition depends explicitly on $v$ and with this gauge we would be computing the effective potential (a gauge-dependent quantity) at different values of $v$ in different gauges. For this reason we will instead use the Coulomb gauge-fixing,
\begin{equation}
\mathcal{L}_{gauge} = -\frac{1}{2\xi}\left(
\nabla.\Ab \right)^2,
\end{equation} followed by the $\xi\rightarrow 0$ limit. For the one-loop calculations described in this paper the use of Coulomb gauge is a modest calculational complication that preempts more complicated conceptual questions.

The quadratic part of the action is given, in momentum space by \begin{align}\label{eq:S_mom}
S_{quad} &=\int \frac{d^4p}{(2\pi)^4}
\begin{pmatrix}
\chi_R(-p) &
\chi_I(-p)  &
A_0(-p) &
A_\parallel(-p)
\end{pmatrix}\times  \nonumber \\
&\phantom{\int \frac{d^{4}p}{(2\pi)^{4}}}\ \begin{pmatrix}
-\frac{p^2}{2M}+\mu     &    ip_0    & Zev    & 0    \\
-ip_0                                & -\frac{p^2}{2M}+\mu      &     0     &   \frac{iZev p}{2M}     \\
Zev                                    &   0                                    & \frac{p^2+\Pi}{2}     &   -\frac{p_0}{2p}(p^2+\Pi)\\
0                                   &   -  \frac{iZev p}{2M}     &  -\frac{p_0}{2p}(p^2+\Pi)   &   \frac{1}{2}\left (  -\frac{Z^{2}e^2v^2}{M}+p_0^2 - \frac{p^2}{\xi}+\frac{p_0^2 \Pi}{p^2} \right)
\end{pmatrix}
\begin{pmatrix}
\chi_R(p)\\
\chi_I(p)\\
A_0(p)\\
A_\parallel(p)
\end{pmatrix}.
\end{align} 
The eigenvalues of the matrix in \eqref{eq:S_mom} are $p_0-iE_p$ and $p_0+iE_p$, where
\begin{align}\label{eq:Ep}
		E_{p}^{2}	=&\	\frac{p^{4}-2M p^{2}\mu -2 M m_{A}^{2}\mu \xi}{p^{4}+4 m_{A}^{2}p^{2}\xi-2 M m_{A}^{2}\mu\xi}\left(	\frac{p^{2}}{2M}\left(	\frac{p^{2}}{2M}-\mu	\right)	+\frac{p^{2}m_{A}^{2}}{p^{2}+\Pi}\right)	\nn\\
		\stackrel{\xi\rightarrow 0}{\longrightarrow}&
		\left(   \frac{p^2}{2M}-\mu\right)^2 +  \left(   \frac{p^2}{2M}-\mu\right) \frac{2M m_A^2}{p^2+\Pi},
\end{align}

where $m_A^2 = 4\pi \alpha Z^2 v^2/M$. Eq.~(\ref{eq:Ep}) gives the dispersion relation for the quasiparticles in the system (after the analytic continuation $p_0\rightarrow i p_0$). At zero temperature and up to higher loop corrections we can use the tree value of the chemical potential $\mu=0$ in the dispersion relation above. In that case, the result in \eqref{eq:Ep} becomes independent of the gauge fixing parameter $\xi$ (as it should since it is an observable quantity) and it agrees with the dispersion relation obtained in using the unitary gauge \cite{Bedaque:2011hs}.

The computation of the one-loop part of the potential $V^{(1)}$ can then proceed in the usual fashion.  We introduce $\delta$ a spurious offset to $p_{0}^{2}+E_{p}^{2}$, do formal manipulation, and then eliminate $\delta$.
\begin{align}
	V^{(1)} &= 	\half T \sum_{p_{0}}\int \frac{d^{3}p}{(2\pi)^{3}} \ln\left(	p_{0}^{2}+E_{p}^{2}	\right) 
			&&= \left.\half T \sum_{p_{0}}\int \frac{d^{3}p}{(2\pi)^{3}} \ln\left(	p_{0}^{2}+E_{p}^{2}	+ \delta \right)\right|_{\delta=0}	\nn\\
			&=	\int^{0}d\delta\ \frac{d}{d\delta}\half T \sum_{p_{0}}\int \frac{d^{3}p}{(2\pi)^{3}} \log\left(	p_{0}^{2}+E_{p}^{2}	+ \delta \right)
			&&=	\half \int^{0}d\delta\ T \sum_{p_{0}}\int \frac{d^{3}p}{(2\pi)^{3}} \oneover{\left(	p_{0}^{2}+E_{p}^{2}	+ \delta \right)}	\nn\\
			&=	\half \int^{0} d\delta\ \int \frac{d^{3}p}{(2\pi)^{3}} \oneover{2 \sqrt{E_{p}^{2}+\delta}}\coth\left(	\half\beta\sqrt{E_{p}^{2}+\delta}	\right)
			&&=	T \int \frac{d^{3}p}{(2\pi)^{3}} \ln 2 \sinh\left(	\half \beta E_{p}	\right)												\nn\\
			&= \int \frac{d^{3}p}{(2\pi)^{3}} \half E_{p} + T \int \frac{d^{3}p}{(2\pi)^{3}} \ln\left(	1-e^{-\beta E_{p}}	\right),
			\label{eq:VEp}
\end{align}
where the sum is over the discrete values $p_0=2\pi T j$, $j=0, \pm 1, \cdots$. Had we kept the full $\Pi(p_0,\mathbf{p})$ instead of merely $\Pi=m_s^2$ we would have encountered cuts in the complex $p_0$ plane and the effective potential would have a more complicated form.  We will ignore the temperature-independent piece when exploring this effective potential in pursuit of its corresponding phase diagram.
  
\section{Phase diagram}\label{sec:phase-diagram}
The minimization of $V$ in relation to $v$ gives us the actual expectation value of the condensate at any given value of the chemical potential $\mu$. We would like, however, to have the ion density $n$ fixed in order to neutralize the charge of the electrons.  We then have to solve simultaneously the pair of equations
\begin{subequations}\label{eq:dV}
	\begin{align}
	n &= 	-\frac{\partial V}{\partial\mu} = v^2+
		 	\int \frac{d^3p}{(2\pi)^3}
		 	\frac{1}{e^{\beta E_p}-1}\frac{1}{E_p} \left[  \frac{p^2}{2M}-\mu+ \frac{M m_A^2}{p^2+m_s^2}  \right],
		 	\label{eq:dVdmu}\\
	 0&=	\phantom{-}\frac{\partial V}{\partial v} = -\mu v+v  \int \frac{d^3p}{(2\pi)^3}
	 		\frac{1}{e^{\beta E_p}-1}\frac{1}{E_p} \left( \frac{p^2}{2M}-\mu\right) \frac{4\pi Z^{2} \alpha}{p^2+m_s^2} 
			\label{eq:dVdv}   .
	\end{align}
\end{subequations}

The first terms in these equations come from the tree level contribution $V^{(0)}=-\mu v^2$ to the effective potential. In the absence of dynamical electrons and screening effects ($m_s=0$), \eqref{eq:dV} were derived in \reference{Fetter1970464}.

We notice that the dispersion relation dependence on $\mu$, shown in equation \eqref{eq:Ep}, is important in deriving these relations, even if $\mu$ is set to zero afterwards. The non-relativistic limit of the dispersion relation obtained in \reference{Rosen:2010es}, for instance, differs from ours at finite values of $\mu$ and is, consequently, at odds with  \reference{Fetter1970464}.

\begin{figure}[tbp]
\centerline{ {\epsfxsize=3.0in\epsfbox{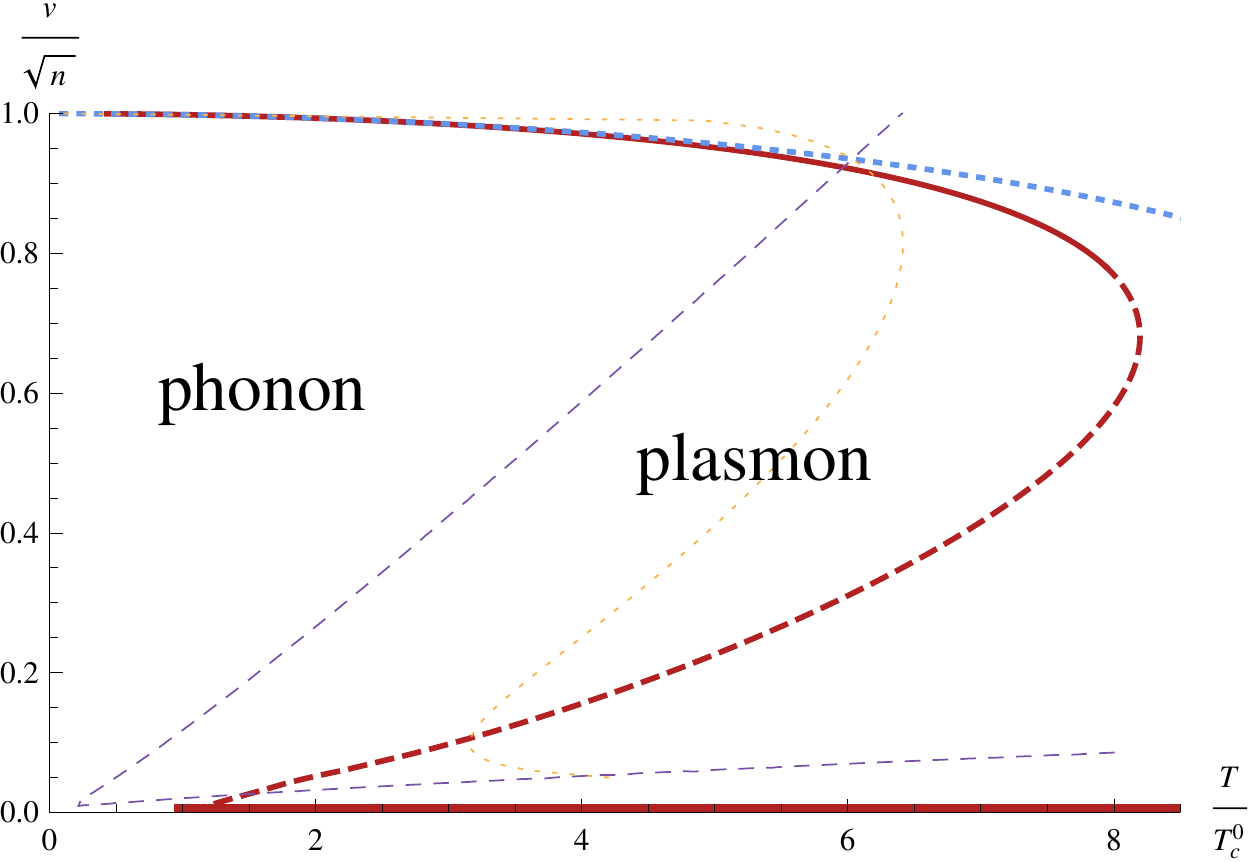}} }
 \noindent
  \caption{The red curve is the solution to \eqref{eq:dVdmu0}; it is shown as dashed where the condition $\mu \ll p^2/2M$ is violated and should not be trusted. The result of the phonon (dotted blue) and plasmon (dotted orange) approximations are also shown. The purple dashed line separates the phonon-dominated from the plasmon-dominated regions. }
\label{fig:phase_diagram}
\end{figure}  

To the extent that higher loop contributions to the effective potential are small \eqref{eq:dV} determine the variation of the condensate $v=v(T)$ with the temperature. 
However,  as we can see from \eqref{eq:dVdv}, $\mu$ cannot be negative if $v\neq 0$. We are led then to expect that the condensate forms at positive values of $\mu$. However, at positive $\mu$, $E_{p}^{2}$ can be negative for some values of $p$, and thus the effective potential is complex at positive $\mu$.  A complex one-loop effective potential is a common occurrence and signals an instability \cite{Weinberg:1987vp}. Frequently, this instability is an artifact of the loop expansion and can be cured by a resummation of higher loop contributions, as it occurs, for instance, in the relativistic $\lambda\phi^4$ model at finite temperature \cite{Dolan:1973qd}.  We are unable at this point to identify the necessary resummations needed in our model. But, just like in the models discussed in \reference{Dolan:1973qd}, the un-resummed one-loop potential already carries important information about the thermodynamics of our problem. We will now proceed to extract as much information from the one-loop effective potential as possible.

The analysis of \eqref{eq:dV}  is simple in the  $v=0$ case and found in textbooks. In this case equation \eqref{eq:dVdmu} becomes
\begin{equation}
n =
 \int \frac{d^3p}{(2\pi)^3}
 \frac{1}{e^{\beta (\frac{p^2}{2M}-\mu)}-1}.
\end{equation} Following the usual analysis, $\mu$ is negative for 
\begin{equation}
T>T^{(0)}_c=\frac{2\pi}{M} \left(\frac{n}{\zeta(3/2)}\right)^{2/3} = \left(  \frac{9\pi}{2\zeta(3/2)^2} \right)^{1/3}\frac{1}{M l^2},
\end{equation} and vanishes at $T=T^{(0)}_c$. For $T<T^{(0)}_c$ it is impossible to satisfy the equations, which signals the need for a non-vanishing condensate. The line $v(T)=0$ is shown in  \figref{fig:phase_diagram}.

In order to consider non-zero values of $v$ our strategy will be to neglect $\mu$ as compared to $p^2/2M$ in \eqref{eq:dV} obtaining
\begin{subequations}\label{eq:dV0}
	\begin{align}
	n -v^2&= 
	 \int \frac{d^3p}{(2\pi)^3}
	 \frac{1}{e^{\beta E_p}-1}\frac{1}{E_p} \left[  \frac{p^2}{2M}+ \frac{M m_A^2}{p^2+m_s^2}  \right],
	 \label{eq:dVdmu0}\\
	 \mu&=  \int \frac{d^3p}{(2\pi)^3}
	 \frac{1}{e^{\beta E_p}-1}\frac{1}{E_p} \frac{p^2}{2M} \frac{4\pi Z^{2}\alpha}{p^2+m_s^2}
	 \label{eq:dVdv0} ,
	\end{align} 	
\end{subequations}
with $E^2_p=(p^2/2M)^2+p^2 m_A^2/(p^2+m_s^2)$, which obviates the complex effective potential problem. We then solve \eqref{eq:dV0} and carefully verify the regime of validity of the $\mu\ll p^2/2M$ approximation. We will solve the equations both numerically and, in some limits, analytically. The numerical solution of \eqref{eq:dVdmu0}  and $a_0/l=35$ (corresponding to a density of $\rho=4.6\times10^5 g/cm^3$ is shown as the thick line in \figref{fig:phase_diagram}. 

\figref{fig:phase_diagram} has the typical shape of a first order phase transition. In the usual scenario, for values of temperatures where three solutions exist ($T_c^{(0)} < T \alt 8T_c^{(0)}$ in \figref{fig:phase_diagram} ), two are locally stable and the middle one is unstable. The $v=0$ solution is stable for temperatures higher than a critical value $T_c > T_c^{(0)}$  and the $v\neq 0$ is only metastable. At temperatures lower than $T_c$ the roles between the $v=0$ and $v\neq 0$ are reversed. Thus, the condensate $v$ jumps discontinuously to zero as the temperature is increased past $T=T_c$. As we will see in the next section, the curve shown in figure \figref{fig:phase_diagram} cannot be trusted for all values of $v$ and the situation in our model is more complicated.

\begin{figure}[tbp]
\centerline{ {\epsfxsize=3.0in\epsfbox{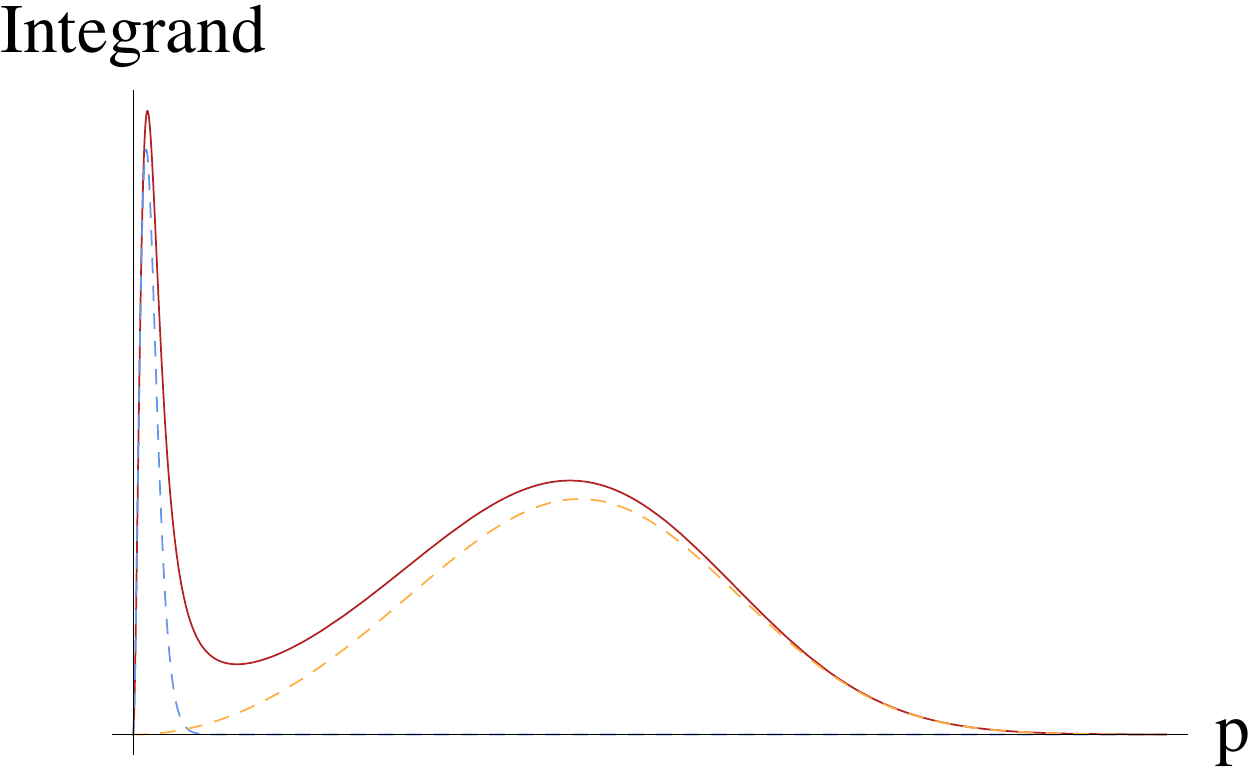}} }
 \noindent
 \caption{Integrand in the second equation in \eqref{eq:dVdmu0} as a function of momentum (full line). The dotted lines correspond to the phonon and plasmon approximations. }
\label{fig:humps}
\end{figure}

We now assess the validity of the approximation $\mu \ll p^2/2M$ leading to our values of $v(T)$. For any given value of $v$ and $T$ we can estimate $\mu$ by using \eqref{eq:dVdv0}. This estimate will be accurate if $\mu$ is indeed negligible compared to $p^2/2M$ but not otherwise. In this sense, the use of \eqref{eq:dVdv0} conservatively estimates the range of validity of the $\mu \ll p^2/2M$ approximation.
The estimate of $p^2/2M$ is a little trickier. Ordinarily, the value of $p^2/2M$ could be estimated from the knowledge of the typical value of $p$ contributing to the integral in \eqref{eq:dVdmu0}. The integrand in \eqref{eq:dVdmu0}, however, has a double hump structure dominated by two widely separate scales as shown in  \figref{fig:humps}. Depending on the values of $v,T$, one or the other hump will dominate the value of the integral. Fortunately, analytical approximations are available in these two cases.

For the lower momentum hump we are in the ``phonon region" where the  approximations 
\begin{subequations}\label{eq:phonon_cond}
	\begin{align}
	 	p^2  &\ll m_s^2,\\
	 	\left(  \frac{p^2}{2M} \right)^2&\ll \frac{m_A^2 p^2}{m_s^2} 		
	\end{align}
\end{subequations}
are adequate. In this region the dispersion relation approaches that of a phonon \cite{Bedaque:2011hs} 
\begin{equation}
	E_p\approx \frac{m_A}{m_s} p
\end{equation} 
and the integrals in \eqref{eq:dV0} can be analytically calculated:
\begin{subequations}\label{eq:dV_phonon}
	\begin{align}
		n&=v^2+\frac{M m_s}{12 m_A}T^2 \label{eq:dVdmu_phonon}, \\
		\mu &= \frac{Z^{2}\pi^3}{15}\frac{\alpha m_s^3}{M}\frac{T^4}{m_A^5} \label{eq:dVdv_phonon}.
	\end{align} 
\end{subequations}
Equation \eqref{eq:dVdmu_phonon} determines the condensate $v(T)$ as a function of $T$.
Its value is plotted as the (blue) dotted line in \figref{fig:phase_diagram}.

\begin{figure}[tbp]
\centerline{ {\epsfxsize=3.0in\epsfbox{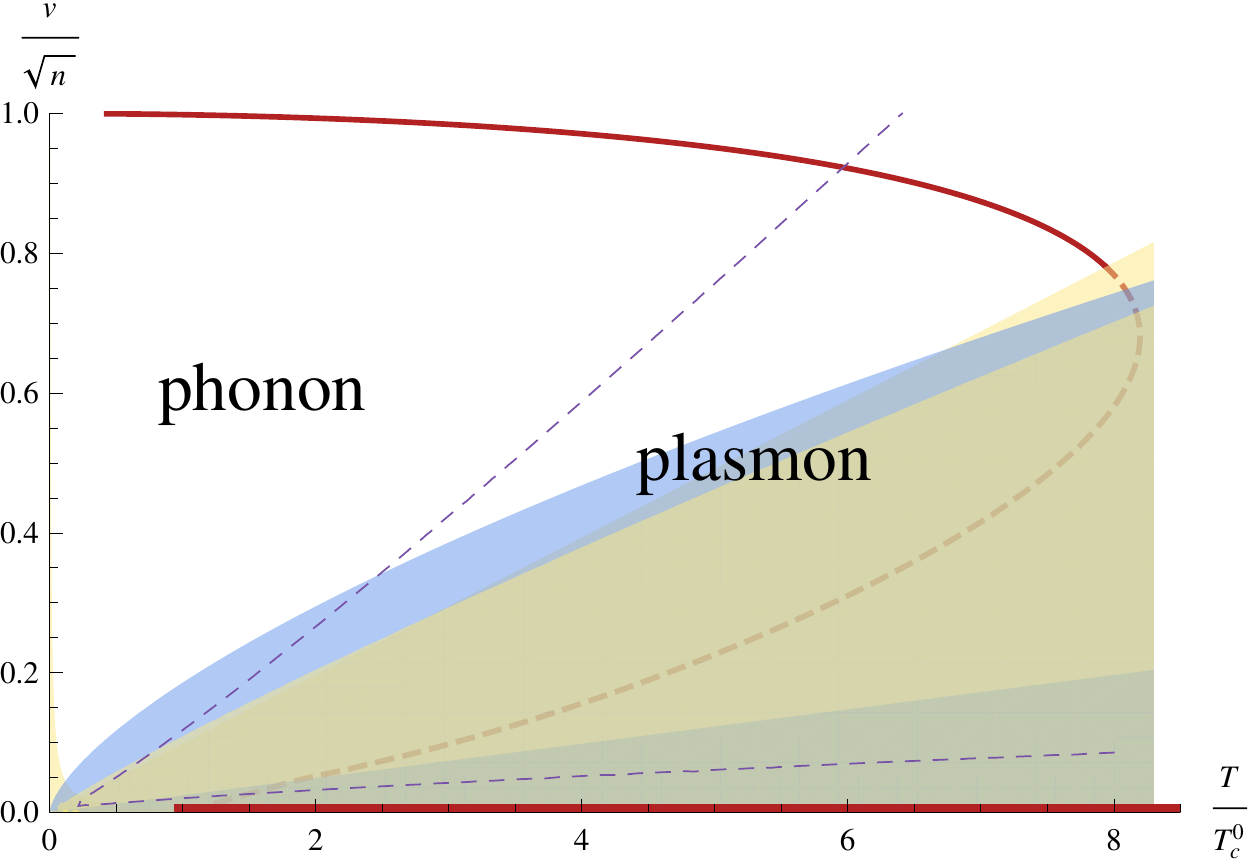}} }
 \noindent
 \caption{Regions where the phonon (plasmon) approximation is not valid are shown in blue (yellow). The purple dashed line separates the phonon-dominated from the plasmon-dominated regions. The red curve is the solution to \eqref{eq:dVdmu0}; it is shown as dashed where the condition $\mu \ll p^2/2M$ is violated and should not be trusted.}
\label{fig:exclude}
\end{figure}

For the phonon approximation to \eqref{eq:dV0} to be legitimate it is necessary that the equations \eqref{eq:phonon_cond} be satisfied. 
The integrals in \eqref{eq:dV0} are cutoff by the Boltzman factor $e^{-m_A p/(T m_s)}$. So, the typical value of the momentum is $p\approx m_s T/m_A$. Using this value of $p$, both conditions in \eqref{eq:phonon_cond} become
\begin{equation}
T \ll m_A.
\end{equation} 
Since $v^2<n$, the second condition in  \eqref{eq:phonon_cond}  follows from the first one. The region excluded by this condition  is shown as the darker blue area in \figref{fig:exclude}.  We can now compare $\mu$ to $p^2/2M$. We find
\begin{equation}\label{eq:muless_phonon}
\mu \ll \frac{p^2}{2M} \Rightarrow \frac{Z^{2}\pi^3}{15} \alpha m_s \frac{T^2}{m_A^3} \ll 1.
\end{equation} 
The region where equation \eqref{eq:muless_phonon} fails is shown in blue in \figref{fig:exclude}.

This condition $T \ll m_A$ is always satisfied when \eqref{eq:muless_phonon} holds and for the relevant values of the density parameter $l/a_0$. We conclude then that, as long as the phonon region dominates the integrals in \eqref{eq:dVdmu0}, we are justified in neglecting $\mu$.

At higher momentum, on the second hump of the integrand, the ``plasmon approximation" is adequate:
\begin{subequations}\label{eq:plasmon_cond}
	\begin{align}
		  m_{s}^{2}		&\ll	p^2						\\
		 \frac{p^2}{2M} &\ll 	\frac{M m_A^2}{p^2} .
	\end{align}
\end{subequations}
In this region, the dispersion relation becomes that of a massive excitation (plasmon)
\begin{equation}
	E_p = \sqrt{\left(  \frac{p^2}{2M} \right)^2+  m_A^2},
\end{equation}   
the integrals are cutoff by the exponential statistical factor $e^{-\sqrt{(p^2/2M)^2+m_A^2}/T} \approx e^{-m_A/T}  e^{-p^4/8M^2 m_A T}$. Also it turns out, \eqref{eq:plasmon_cond} implies $\Rightarrow T \ll \frac{1}{4} m_A$ and hence we can drop 1 compared to $e^{\beta E_p}$ in the integrand. The typical momentum is given by
\begin{equation}
	p^2 \approx \sqrt{8 m_A T} M. 
\end{equation}
With the approximations in \eqref{eq:plasmon_cond}, \eqref{eq:dV0} reduces to
\begin{subequations}\label{eq:dV_plasmon}
	\begin{align}
		n 	&= 	v^2+\frac{\Gamma(5/4)}{2^{1/4} \pi^2} M^{3/2} m_A^{5/4} T^{1/4} e^{-m_A/T},
				\label{eq:dVdmu_plasmon}\\
		\mu	&= 	\frac{2^{1/4} Z^2 \Gamma(3/4)}{\pi} Z^2 \alpha \sqrt{M}\frac{T^{3/4}}{m_A^{1/4}} e^{-m_A/T}
				\label{eq:dVdv_plasmon}		
	\end{align}
\end{subequations}
We can now verify for which values of $v$ and $T$ the plasmon approximation is valid:
\begin{subequations}
\begin{align} 
		m_s^2  			&\ll	p^2 					&\Rightarrow&	&\frac{m_s^4}{8 M^2 m_A^2} 	&\ll T,								\label{eq:plasmon_validity_ms}\\
		\frac{p^2}{2M}  &\ll	\frac{M m_A^2}{p^2}  	&\Rightarrow&	&T 							&\ll \frac{1}{4} m_A,				\label{eq:plasmon_validity_ma}\\
		\mu 			&\ll	\frac{p^2}{2M} 			&\Rightarrow&	&\frac{Z^2\Gamma(3/4)}{2^{1/4}\pi}	\frac{\sqrt{M}T^{1/4}}{m_A^{3/4}} \alpha\  e^{-m_A/T} &\ll 1.		\label{eq:plasmon_validity_mu}
\end{align}
\end{subequations}
The first condition \eqref{eq:plasmon_validity_ms} excludes a tiny region of the $v-T$ plane near $T\approx 0$ where our calculations which neglect the temperature independent one-loop effective potential are not valid.  The second and third condition are actually very similar since the exponential factor $e^{-m_A/T}$ is very small when as $T\alt m_A$. The areas excluded by these conditions  are shown in yellow in \figref{fig:exclude}.

Finally, the phonon and plasmon region contributions to the integral in \eqref{eq:dVdmu0} should be compared by taking the ratio of \eqref{eq:dVdmu_phonon} and \eqref{eq:dVdmu_plasmon}. We find 
\begin{equation}\label{eq:comparison}
\frac{\left.\left(	n-v^2	\right)\right|_{\text{phonon}}}{\left.\left(	n-v^2	\right)\right|_{\text{plasmon}}} \sim \frac{2^{1/4} \pi^2}{12\ \Gamma(5/4)}\frac{m_s T^{7/4}}{\sqrt{M} m_A^{9/4}}   e^{m_A/T}.
\end{equation}  
The contour separating the phonon-dominated from the plasmon-dominated regions is shown as a dashed line \figref{fig:phase_diagram} and \figref{fig:exclude}. 
For $v\approx \sqrt{n}$, the only part of the $v=v(T)$ solution where those approximations make sense, the phonon contribution dominates at small temperatures while the plasmon contribution dominates at higher temperatures. 

We have now our full verification of the $\mu=0$ approximation. On the upper branch of the $v=v(T)$ solution up to  $T\alt 5 T_c^{(0)}$ the phonon and the $\mu \approx 0$ approximations are valid. At higher $T$ until $T\approx 7.5 T_c^{(0)}$, the plasmon and the $\mu \approx 0$ approximations are valid. Beyond that and for the lower branch of $v=v(T)\alt 0.7 \sqrt{n}$ our approximations, including the neglect of $\mu$, are no longer valid. For this reason we draw the $v=v(T)$ solution as a dashed line in \figref{fig:phase_diagram} and \figref{fig:exclude}. 

The numerical examples presented in the figures correspond to a fixed value of the density parameter $l/a_0 = 1/35$. It turns out that the dependence of $v/\sqrt{n}$ and other quantities on the density is very mild, particularly if we remember that only a relatively narrow range of densities around $10^5\ g/cm^3$ is phenomenologically relevant. In fact, in addition to the dependence of $m_s \sim k_F^{1/2} \sim n^{1/6}$, only the overall normalization of the integral on \eqref{eq:dVdmu0} is dependent on $n$. As a rule, however, there is a reduction of the phonon dominated region in the phase diagram as the density is raised, as can be seen on \eqref{eq:comparison}.
  
\section{Global stability and a conjecture}\label{sec:global}
Having identified the locally stable states of the model we now study their global stability. 
In a certain range of temperatures two states, the trivial $v=0$ and the condensed $v\neq 0$ one satisfy the conditions for the local minimization of the effective potential. We want now to decide which one is the global minimum. There two states, however, are obtained from solving \eqref{eq:dVdmu0} and, consequently, have the same particle density $n$. In order to decide which one is the stable state we have to compare their free energies defined by
\begin{equation}
F(n,T) = V_{eff}(v,T)+\mu n,
\end{equation} 
where $\mu$ and $v$ are the solutions to \eqref{eq:dVdmu0} and \eqref{eq:dVdv0} for  given values of $T$ and $n$. The evaluation of $F$ is particularly simple right at $T=T_c^{(0)}$. 

Let us first compute this value for the trivial $v=0$ solution. Right at $T=T_{c}^{0}$ the uncondensed solution has $\mu=0$, and we find
\begin{align}
F_{v=0}\left(	n, T_{c}^{(0)}	\right) &=			T_{c}^{(0)}\int \frac{d^3p}{(2\pi)^3} \ln\left(  1-e^{-\frac{p^2}{2MT_c^{(0)}}}\right)	\nn\\
		&= 		-\zeta(5/2)\ T_c^{(0)}\ \left( \frac{MT_c^{(0)}}{2\pi} \right)^{3/2}				\nn\\
		&=		-\frac{3^{5/3}\zeta(5/2)}{2^{7/3}\zeta^{5/3}(3/2)}    \frac{1}{Ml^5}
\end{align} 
This is to be compared to the free energy of the $v\neq 0$ states. The $T=T_c^{(0)}$, $v\neq 0$ solution is well within the phonon dominated region where the free energy can be computed, with the results in \eqref{eq:dV_phonon}, to be
\begin{align}
F_{\text{phonon}}\left(	n, T	\right) &=
\mu(n-v^2)+T\int \frac{d^3p}{(2\pi)^3} \ln\left(  1-e^{-\frac{m_A p}{m_sT}}\right)\nn\\
&=
\frac{\pi^3 Z^{2} \alpha m_s^4 T^6}{180 m_A^6} - \frac{\pi^2 m_s^3 T^4}{90 m_A^3}\\
F_{\text{phonon}}(n,T_{c}^{(0)})&=
 - \frac{3^{1/6}\pi^{7/3}}{5\ 2^{1/3}\zeta^{8/3}(3/2)Z^{5/2}} \oneover{M l^{5}}\left(	\frac{m}{M}	\right)^{3/2}\left(	1 - \frac{3^{1/6}\pi^{4/3}}{2 \zeta^{4/3}(3/2)Z^{5/6}}\sqrt{\frac{m}{M}}	\right)
\end{align} 
The free energy of the $v=0$ solution is more negative than the free energy of the condensed state by a factor proportional to the large parameter $(M/m)^{3/2}$.  This shows that at all temperatures where the $v=0$ state exists it is it, and not the condensed $v\neq 0$ state, which is the globally stable state of the system. A numerical calculation of the free energy using the solution of \eqref{eq:dVdmu0} and \eqref{eq:dVdv0} confirms this result and shows that the difference in free energy between the two competing states {\it increases} as the temperature is further increased past $T=T_c^{(0)}$.

This result is at odds with the standard picture of a first order transition. In the usual case, one has $F_{\text{phonon}} < F_{v=0}$ for all $T<T_c$ where the critical temperature $T_c$ is greater than $T_c^{(0)}$.  Even though we were unable to compute $F(v,n,T)$ for arbitrary $v$, it is easy to see that no function $F(v,n,T)$ could have a set of local and global minima as shown in  \figref{fig:phase_diagram} while also globally favoring the $v=0$ state all the way down to $T_{c}^{(0)}$.  To understand this impossibility, consider the shape of this purported function around $T\approx T_c^{(0)}$ as $T$ is increased. A new minimum at $v=0$ is supposed to appear in addition to the non-trivial one with $v\neq 0$ and immediately become the global minimum of the function. This is clearly impossible.

The only way out of this inconsistency is to assume that the state with small $v$ beats the superconducting state $v\approx\sqrt{n}$ at temperatures smaller than $T_c$. This can occur if the
actual curve $v=v(T)$ has the shape shown in the left panel of \figref{fig:conjecture}. In that case a first order transition occurs at a lower smaller than $T_c^{(0)}$ followed by a second order transition. The shape of the free energy as a function of $V$ at different temperatures indicated on the right panel is sketched on the right panel of \figref{fig:conjecture}. The extra knee in the $v=v(T)$ curve can occur in the region of the $v-T$ plane where our calculation is not under control. The part of our calculation that is under good theoretical control forces us to believe in this more exotic possibility

\begin{figure}
\begin{tabular}{cc}
{\label{fig:conjecture}
\includegraphics[width=0.3\textwidth]{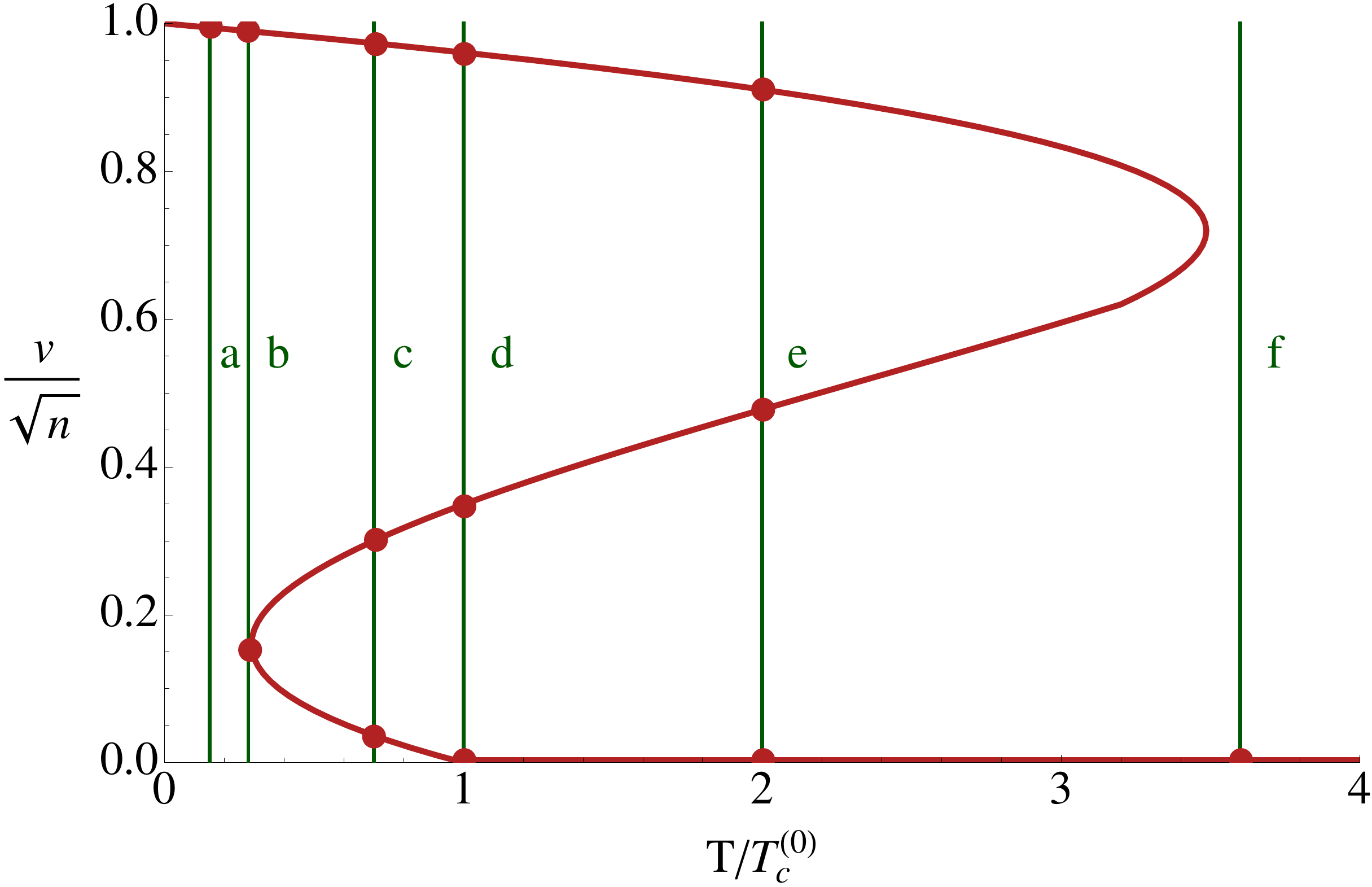}}
&
\begin{tabular}{ccc}
\subfloat[]
{\label{fig:f1}
\includegraphics[width=0.2\textwidth]{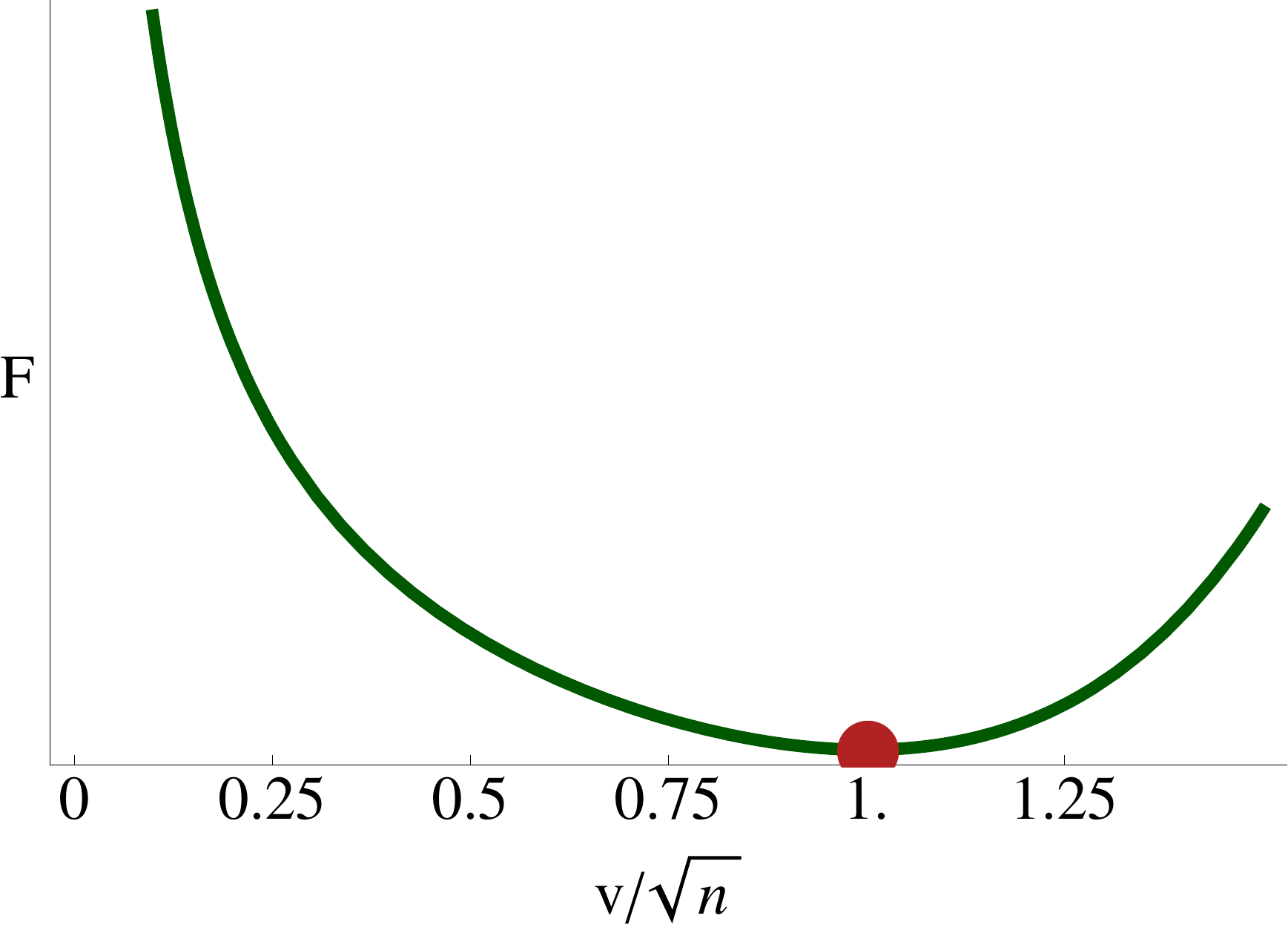}}
&
\subfloat[]
{\label{fig:f2}
\includegraphics[width=0.2\textwidth]{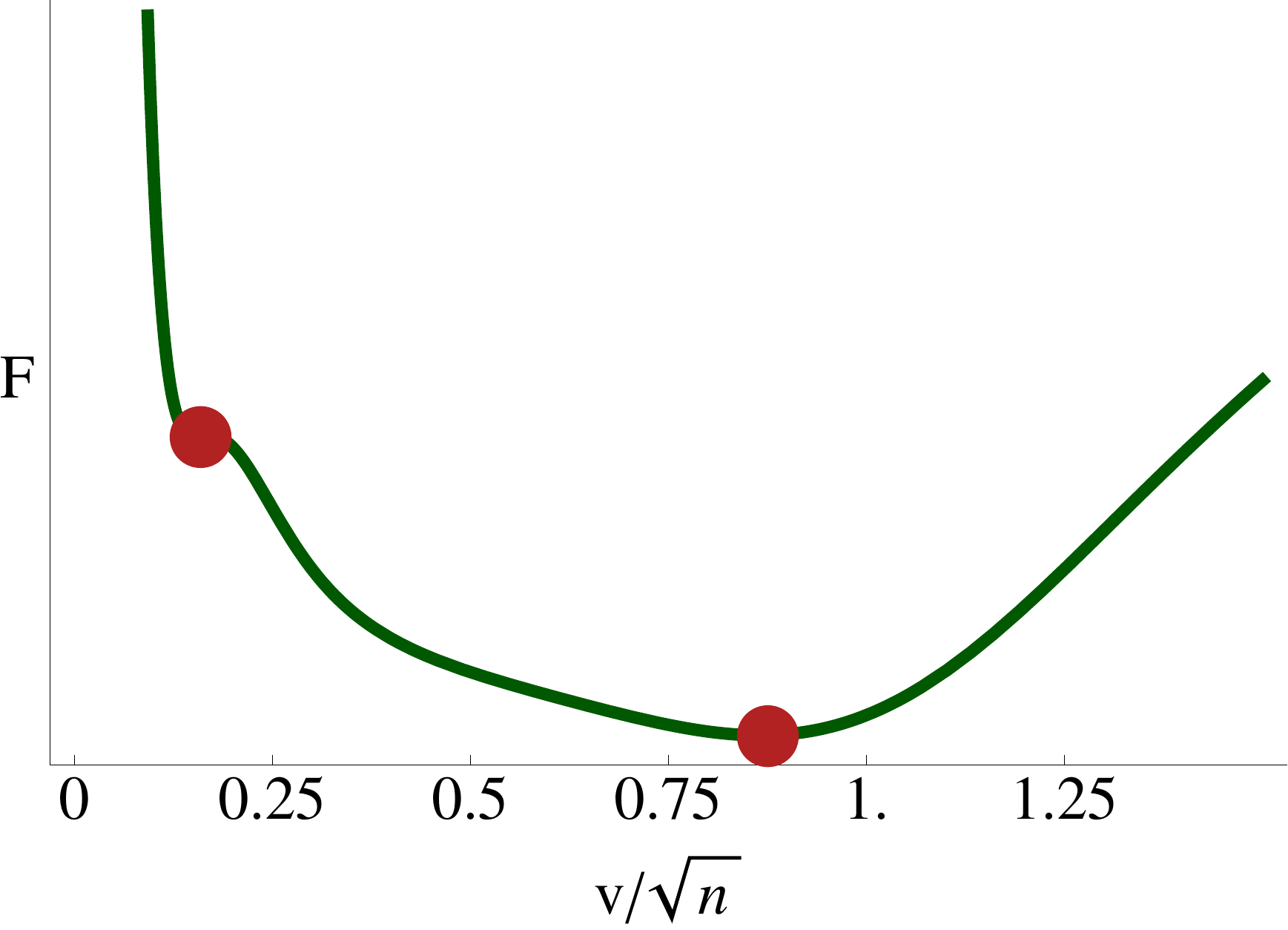}}
&
\subfloat[]
{\label{fig:f3}
\includegraphics[width=0.2\textwidth]{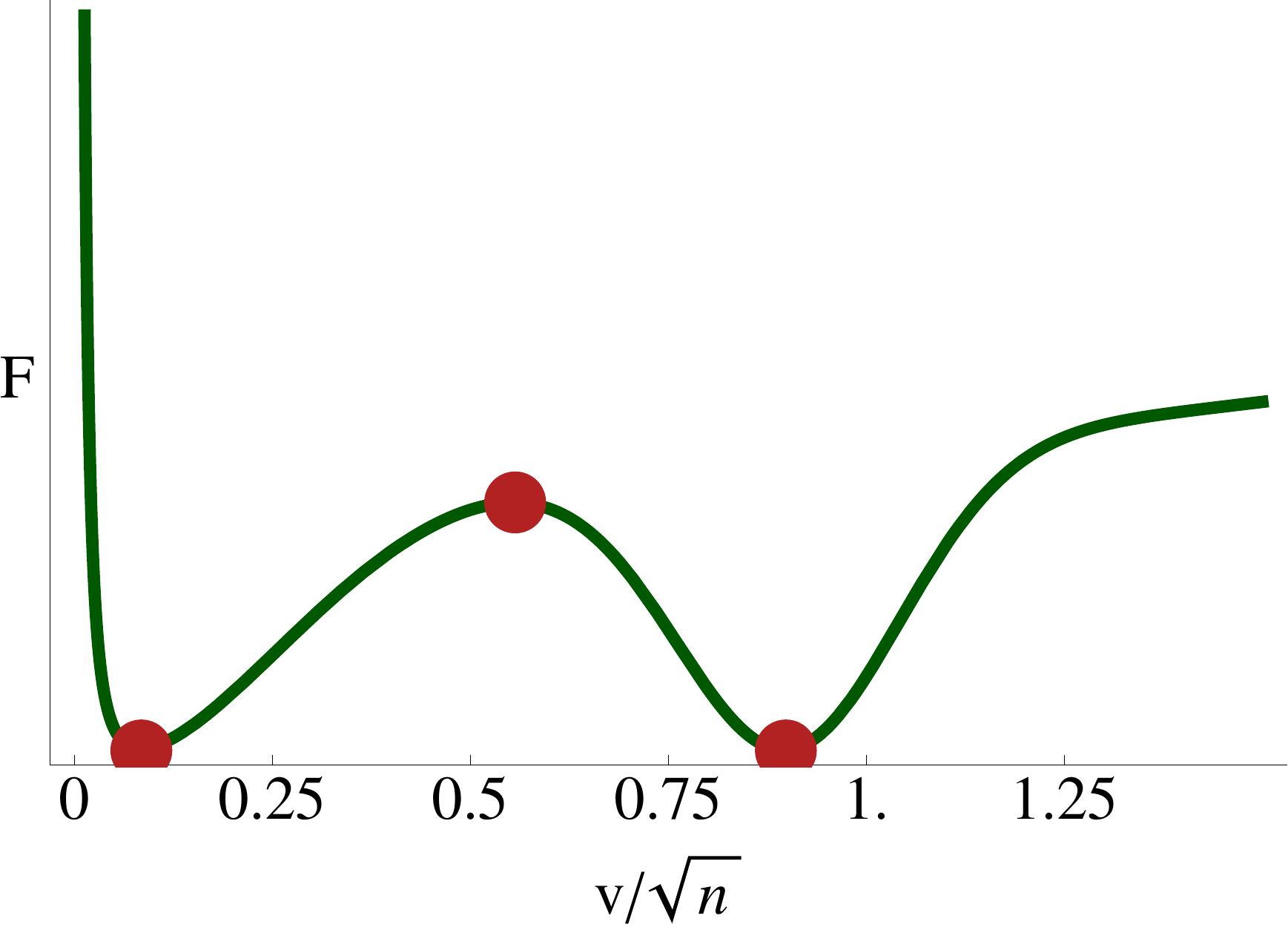}}
\\
\subfloat[]
{\label{fig:f4}
\includegraphics[width=0.2\textwidth]{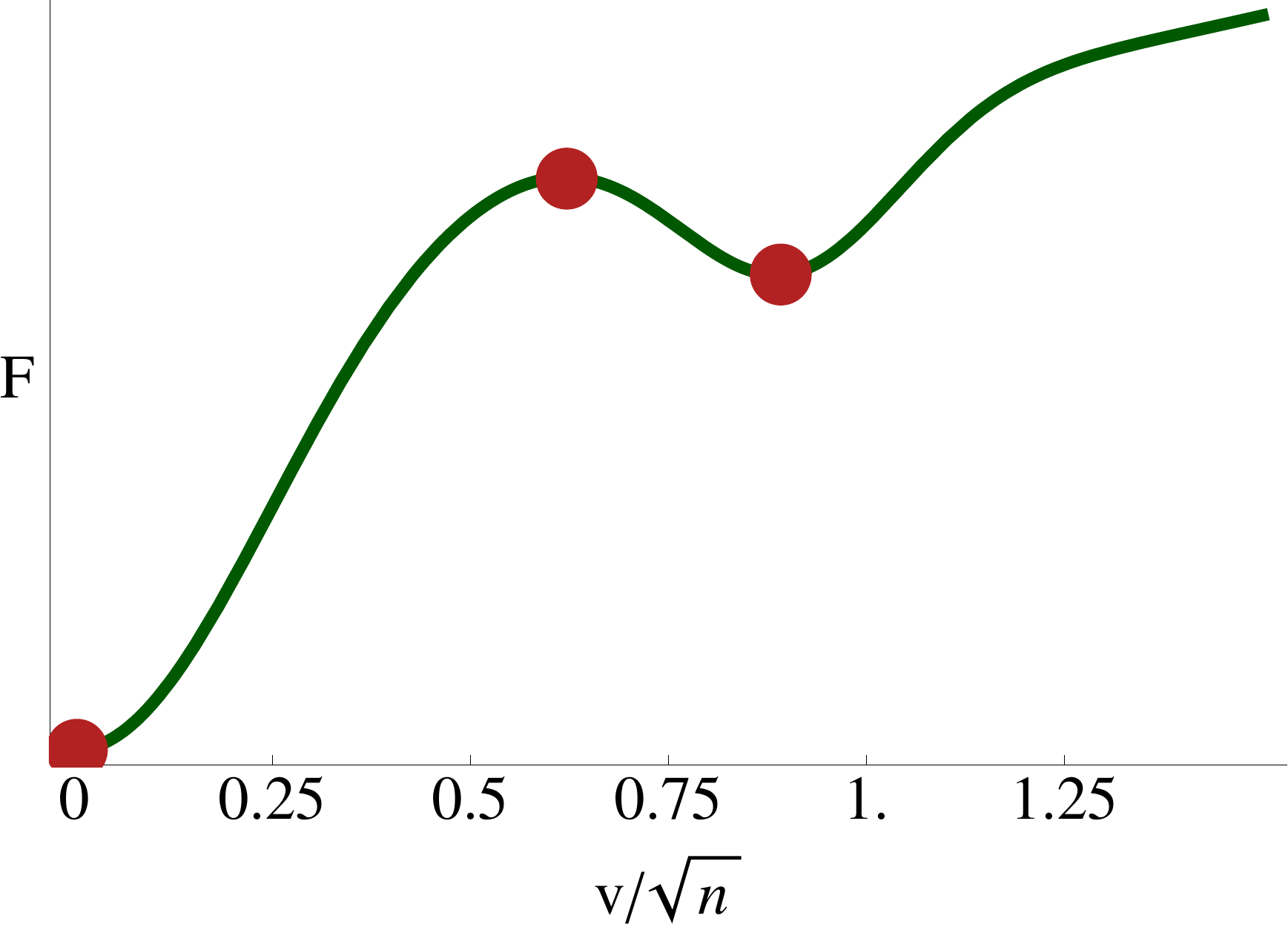}}
&
\subfloat[]
{\label{fig:f5}
\includegraphics[width=0.2\textwidth]{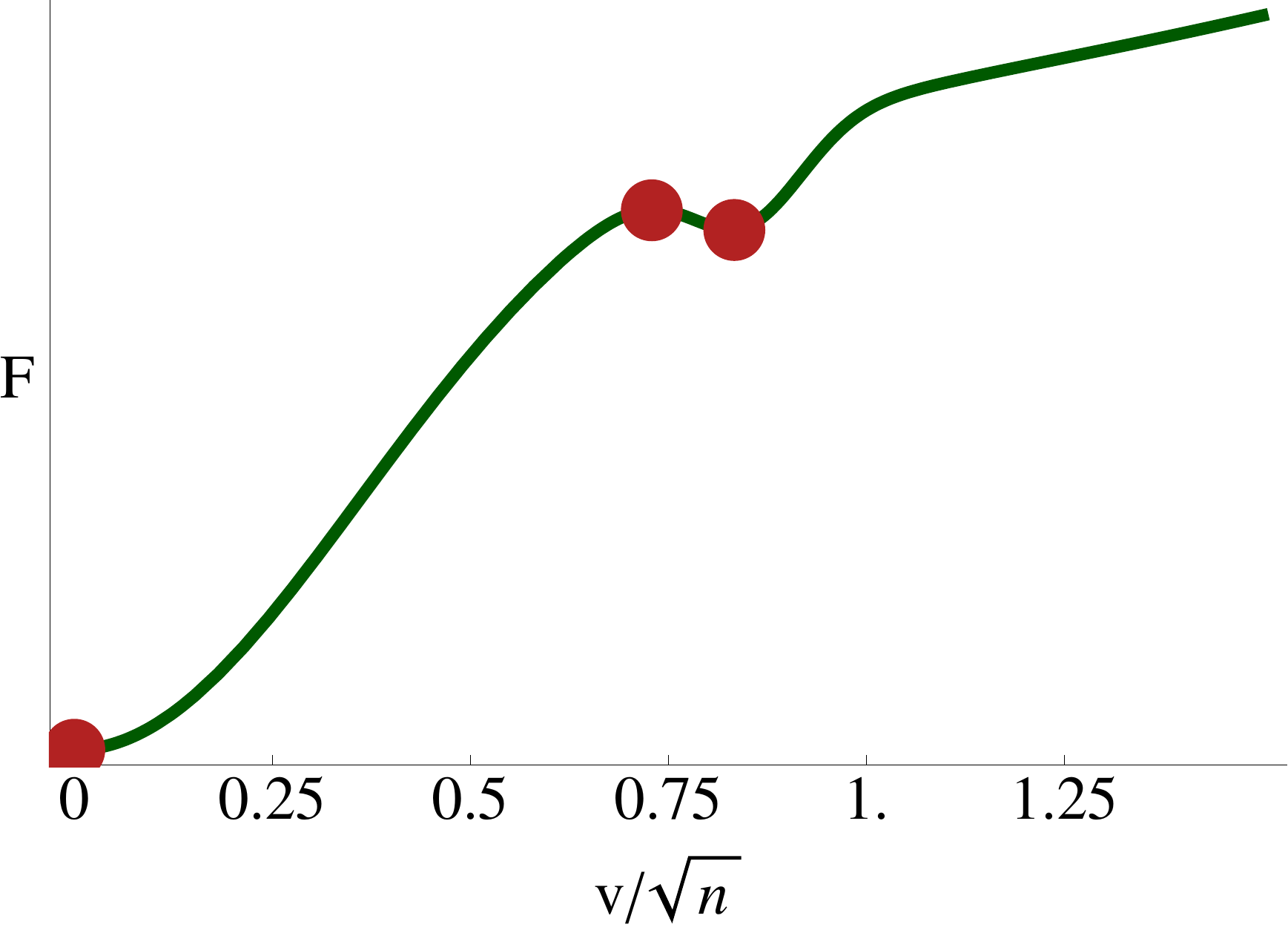}}
&
\subfloat[]
{\label{fig:f6}
\includegraphics[width=0.2\textwidth]{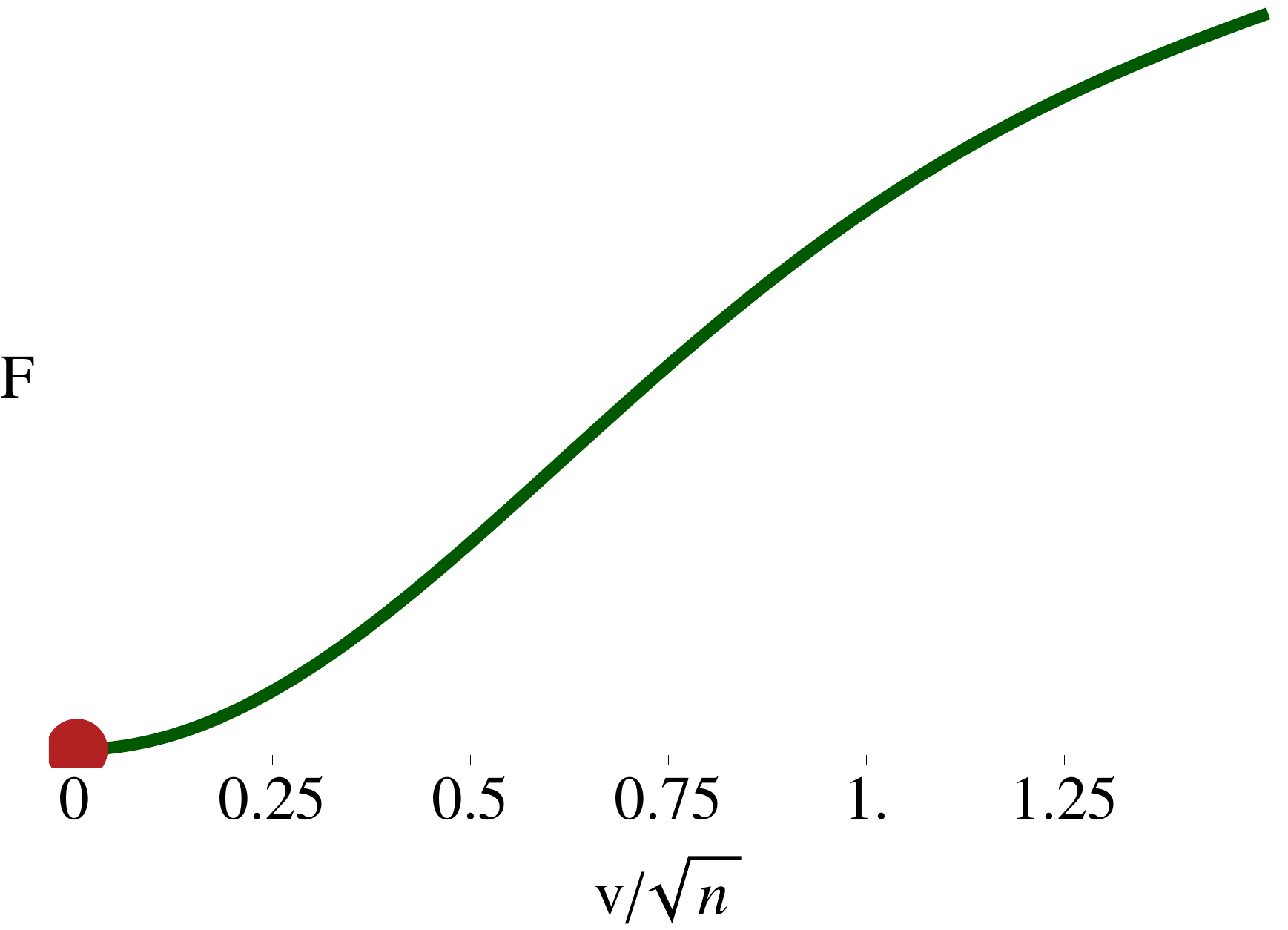}}
\end{tabular}
\end{tabular}
\caption{\label{fig:multi_panel} 
The left panel shows a schematic of our conjectured phase diagram for the condensate $v$ as a function of $T$.  The red curve in the left panel shows the minima in $F(v,n,T)$ for a fixed $n$. The green curves in the right panel are cross-sections of $F(v,n,T)$ at fixed $T$, with the sub-figures in the right panel corresponding to the slices in the left panel with the same label.  At temperature a only the condensed solution exists.  At temperature b the free energy develops an inflection point which turns into one minima and one maxima.  The first-order phase transition happens at temperature c when the wells become equally deep.  By temperature d (ie. $T_{c}^{(0)}$), the well with small $v$ is already favored, and there is a second-order transition to $v=0$.  At higher temperatures the $v\neq0$ minimum vanishes, shown in (e) and (f).
}\label{fig:conjecture}
\end{figure}

\section{Conclusion}\label{sec:conclusion}

Despite the difficulty posed by the fact that the effective potential is complex for positive chemical potential we were able to extract some physical consequences from our one-loop calculation. The main one is the indication of a sequence of a first order transition followed by a second order phase transition. This conclusion was developed by looking at the free energy values computed within the range of validity of the approximations performed and, as such, is quite robust. On the other hand we were not able to compute the critical temperature where the stable (superconducting) and metastable (normal) states trade places. But if the double transition conjecture described above is correct, this temperature is below the free boson critical temperature $T_c^{(0)}$. 

We also found that the  superconducting phase exists, as a metastable state, for temperatures up to about 8 times $T_c^{(0)}$. This conclusion agrees with that in \reference{Rosen:2010es}; this is not entirely a coincidence. Contrary to the present paper, \reference{Rosen:2010es} analyzes the unscreened model ($m_s=0$). But the High $T$ region of the $v=v(T)$ curve is in the plasmon-dominated region where the screening is not important. In addition, the methodological differences between this paper and in \reference{Rosen:2010es} lead to a change in \eqref{eq:dVdmu} that is numerically small and, just like here, the $\mu$ dependence of the dispersion relation is neglected, albeit for different reasons. 

Although the model we analyzed (bosons interacting through a screened Coulomb (Yukawa) potential) is interesting on its own merits, applications to high density physics require a proper treatment of the effects of a dynamical electron background. Technically the main effect is the inclusion of the contribution of the cuts of $\Pi(p_0,\mathbf{p})$ in the effective potential calculation. Physically they correspond to the fact that the actual force between two bosons presents an oscillating component (Friedel oscillations) \cite{Gabadadze:2009zz,Dolgov:2010gy,Gabadadze:2008pj}. A detailed of the influence of these effects on the thermodynamics of the system will be left for a future publication.

We have also not discussed the metastability of the superconducting state in a quantitative fashion. In particular, we have not estimated its lifetime. This is due to the fact that a proper estimate would us require to compute the free energy $F$ as a function $v$ in order to understand the size of the potential barrier separating normal from superconducting phases. Until a deeper understanding of the resummations needed to make sense of the one-loop results is achieved, this calculation is impossible. In fact, all the results discussed here as a well as a confirmation or falsification of our conjectured phase diagram hinge on an understanding of this resummation and that should be viewed as the number one priority for further progress in this topic.

Finally, we can use the results of this paper to assess where the idea of nuclear condensates in dense matter stands. Recall that, for the existence of a intermediate temperature regime where the nuclear condensate can exist it is necessary that the crystallization temperature be smaller than the condensation temperature. If one is willing to consider metastable states, our estimate that the superconducting state extends up to $\approx 8\ T_c^{(0)}$ is similar to the hypothesis made in \cite{Gabadadze:2007si}. A such, the estimate that, at densities around $10^5 g/cm^3$ relevant for white dwarf physics a nuclear condensate should exist, stands unaltered. Only an estimate of the decay time of the false superconducting ground state can decide whether the inclusion of the metastable state is appropriate but, considering the extreme slow evolution of white dwarfs and the fact that they start out at high temperatures, suggest the the metastable state is irrelevant. In that case, only at much higher densities ($\rho \agt 2.4\times 10^7\ g/cm^3$)  and temperatures ($T\agt 10^6\ K$) can the nuclear condensate exist. 

\section*{Acknowledgements}
The authors would like to thank Tom Cohen and Aleksey Cherman for extensive discussions.  This work was supported by the U.S. Dept. of Energy under grant \#DE-DG02-93ER-40762, and E.~B. is also supported by the Jefferson Science Associates under the JSA/JLab Graduate Fellowship program.

\bibliography{nuclear_liquids,astro_refs}
\end{document}